\documentstyle{mn}

\newif\ifAMStwofonts

\ifoldfss
  \ifCUPmtlplainloaded \else
    \NewTextAlphabet{textbfit} {cmbxti10} {}
    \NewTextAlphabet{textbfss} {cmssbx10} {}
    \NewMathAlphabet{mathbfit} {cmbxti10} {} 
    \NewMathAlphabet{mathbfss} {cmssbx10} {} 
  \fi
  \ifAMStwofonts
    \ifCUPmtlplainloaded \else
      \NewSymbolFont{upmath} {eurm10}
      \NewSymbolFont{AMSa} {msam10}
      \NewMathSymbol{\upi}     {0}{upmath}{19}
      \NewMathSymbol{\umu}     {0}{upmath}{16}
      \NewMathSymbol{\upartial}{0}{upmath}{40}
      \NewMathSymbol{\leqslant}{3}{AMSa}{36}
      \NewMathSymbol{\geqslant}{3}{AMSa}{3E}

       \let\le=\leqslant
       
    \fi
  \fi
\fi 

\ifnfssone
  \newmathalphabet{\mathit}
  \addtoversion{normal}{\mathit}{cmr}{m}{it}
  \addtoversion{bold}{\mathit}{cmr}{bx}{it}
  \newmathalphabet{\mathbfit} 
  \addtoversion{normal}{\mathbfit}{cmr}{bx}{it}
  \addtoversion{bold}{\mathbfit}{cmr}{bx}{it}
  \newmathalphabet{\mathbfss} 
  \addtoversion{normal}{\mathbfss}{cmss}{bx}{n}
  \addtoversion{bold}{\mathbfss}{cmss}{bx}{n}
  \ifAMStwofonts
    \ifCUPmtlplainloaded \else
      \UseAMStwoboldmath
      \makeatletter
      \new@mathgroup\upmath@group
      \define@mathgroup\mv@normal\upmath@group{eur}{m}{n}
      \define@mathgroup\mv@bold\upmath@group{eur}{b}{n}
      \edef\UPM{\hexnumber\upmath@group}
      \new@mathgroup\amsa@group
      \define@mathgroup\mv@normal\amsa@group{msa}{m}{n}
      \define@mathgroup\mv@bold\amsa@group{msa}{m}{n}
      \edef\AMSa{\hexnumber\amsa@group}
      \makeatother
      \mathchardef\upi="0\UPM19
      \mathchardef\umu="0\UPM16
      \mathchardef\upartial="0\UPM40
      \mathchardef\leqslant="3\AMSa36
      \mathchardef\geqslant="3\AMSa3E

       \let\le=\leqslant

    \fi
  \fi
\fi 

\ifnfsstwo
  \DeclareMathAlphabet{\mathbfit}{OT1}{cmr}{bx}{it}
  \SetMathAlphabet\mathbfit{bold}{OT1}{cmr}{bx}{it}
  \DeclareMathAlphabet{\mathbfss}{OT1}{cmss}{bx}{n}
  \SetMathAlphabet\mathbfss{bold}{OT1}{cmss}{bx}{n}
  \ifAMStwofonts
    \ifCUPmtlplainloaded \else
      \DeclareSymbolFont{UPM}{U}{eur}{m}{n}
      \SetSymbolFont{UPM}{bold}{U}{eur}{b}{n}
      \DeclareSymbolFont{AMSa}{U}{msa}{m}{n}
      \DeclareMathSymbol{\upi}{0}{UPM}{"19}
      \DeclareMathSymbol{\umu}{0}{UPM}{"16}
      \DeclareMathSymbol{\upartial}{0}{UPM}{"40}
      \DeclareMathSymbol{\leqslant}{3}{AMSa}{"36}
      \DeclareMathSymbol{\geqslant}{3}{AMSa}{"3E}

       \let\le=\leqslant

    \fi
  \fi
\fi 

\ifCUPmtlplainloaded \else
  \ifAMStwofonts \else 
    \def\upi{\pi}
    \def\umu{\mu}
    \def\upartial{\partial}
  \fi
\fi

\title[K-band flux-limited samples of quasars]
{The KX method for producing K-band flux-limited samples of quasars}

\author[S. J. Warren, P. C. Hewett, C. B. Foltz]
       {S. J. Warren,$^1$ P. C. Hewett,$^2$ C. B. Foltz$^3$ \\
       $^1$Blackett Laboratory, Imperial College of Science Technology
       and Medicine, Prince Consort Rd, London SW7 2BZ \\
       $^2$Institute of Astronomy, Madingley Rd, Cambridge CB3 0HA \\
       $^3$MMT Observatory, University of Arizona, Tucson, AZ 85721, USA}

\date{Accepted
      Received
      in original form}

\begin{document}

\maketitle

\begin{abstract}

The longstanding question of the extent to which the quasar population
is affected by dust extinction, within host galaxies or galaxies along
the line of sight, remains open. More generally, the spectral energy
distributions of quasars vary significantly and flux--limited samples
defined at different wavelengths include different quasars. Surveys
employing flux measurements at widely separated wavelengths are
necessary to characterise fully the spectral properties of the quasar
population.  The availability of panoramic near--infrared detectors on
large telescopes provides the opportunity to undertake surveys capable
of establishing the importance of extinction by dust on the observed
population of quasars. We introduce an efficient method for selecting
K--band, flux--limited samples of quasars, termed ``KX'' by analogy
with the UVX method. This method exploits the difference between the
power--law nature of quasar spectra and the convex spectra of stars:
quasars are relatively brighter than stars at both short wavelengths
(the UVX method) and long wavelengths (the KX method). We consider the
feasibility of undertaking a large--area KX survey for damped
Ly$\alpha$ galaxies and gravitational lenses using the planned UKIRT
wide--field near--infrared camera.
\end{abstract}

\begin{keywords}
quasars: general -- quasars: absorption lines -- gravitational lensing
--methods: observational
\end{keywords}

\section{Motivation}

More than three decades since the discovery of quasars the question of
whether our knowledge of the quasar population is affected
significantly by dust extinction, within host galaxies or galaxies
along the line of sight, remains controversial. Obscuration by tori or
other non--spherical components forms a key element of unified schemes
for active galactic nuclei (AGN). Similarly, while estimates of the
extinction optical depth through the discs of spiral galaxies differ,
a number of gravitationally lensed quasars and AGN provide unambiguous
evidence that extinction within the deflector or host galaxies affects
our view of the sources. However, it is not clear if obscuration by
dust merely perturbs our view, increasing the overall quasar space
density, or the frequency of a particular quasar sub--type, by a
factor two or less, or, whether objects in existing quasar surveys
represent only a small fraction of the (unobscured) population.

Much discussion concerning obscured quasars has focussed on whether
such objects, with their spectral energy distributions (SEDs)
steepened by the effects of reddening by dust, would be detectable
using a particular quasar identification technique, such as slitless
spectroscopy or multicolour selection in the optical. In fact, given
that most surveys for quasars rely on flux--limited samples of very
limited dynamic range, the key problem is that quasars will be dimmed
to the extent that they simply do not appear in the sample.
Fortunately, the rapid advance in the fabrication of large
near--infrared detectors means it will soon be viable to undertake
surveys for quasars over large areas of sky that are much less
susceptible to the effects of extinction by dust.

In this paper we introduce a method for selecting samples of quasars,
flux--limited in the K band, that would include quasars whose flux
has been dimmed by dust, and which may have eluded conventional optical
surveys. The method is termed ``KX'', by analogy with the UVX method,
as it similarly exploits the difference between the power--law nature
of quasar spectra and the convex spectra of stars; quasars are brighter
relatively than stars at both short wavelengths (the UVX method) and
long wavelengths (the KX method). In this section we review the
motivation for surveys that can detect reddened quasars. In \S2 we
describe the method. Then, in \S3 we consider the feasibility of
undertaking a large KX survey for damped Ly$\alpha$ galaxies and
gravitational lenses using the planned UKIRT wide--field near--infrared
camera.

\subsection{Quasars and cosmology}

Samples of distant quasars that are bright at optical wavelengths have
proven valuable in a number of areas in extra--Galactic astronomy and
cosmology. Here we focus on the use of quasars for \ i) charting the
history of star formation with lookback time through surveys for
damped Ly$\alpha$ (DLA) galaxies (Pei and Fall 1995), and \ ii)
determining the geometry of the Universe from the measurement of the
frequency of gravitational lensing (Fukugita, Futamase and Kasai 1990,
Turner 1990). The effects of dust in the intervening galaxies are
crucial to the interpretation of such studies. Surveys for quasars at
optical wavelengths are very susceptible to extinction, because the
observed passband corresponds to the rest-frame ultraviolet at
moderate redshift and beyond. Utilising near--infrared wavelengths
produces a significant improvement. For example, a high--redshift
quasar, $z \sim 3$, behind a DLA system at $z=2.5$ with (rest--frame)
$E(B-V)= 0.2\,$mag would suffer $2.8\,$mag extinction in the
observed--frame B--band but only $0.5\,$mag at K. (Except where
otherwise noted all extinction calculations assume an LMC--like
extinction curve below $0.33\mu$ rest--frame (Howarth 1983), the
Mathis (1990) curve at longer wavelengths, and $R_V=3.1$).

\subsubsection{Damped Ly$\alpha$ galaxies and the Universal history of
star formation}

To measure the history of gas consumption by star formation in the
Universe, quasars are used as background sources of light and their
spectra are searched for DLA absorption lines caused by intervening
clouds with a high column density of neutral hydrogen. Bright quasars
are needed to provide sufficiently high signal--to--noise ratio spectra
to measure the absorption line properties of the intervening gas. The
frequency and column densities of the DLA absorption lines yield a
direct measurement of the cosmic density of neutral gas, $\Omega_{\rm
HI}$, (Wolfe 1987). Since dusty DLA absorbers can be missed because the
quasar will be dimmed by extinction below the flux limit chosen for the
spectroscopic observations, the procedure is only valid {\em provided
the obscuring effects of dust are taken into account}. The measured
decline of $\Omega_{\rm HI}$ with time, corrected for the effects of
dust, can be related directly to the history of star formation (under
assumptions about the exchange of gas between the neutral and ionised
gas phases).

Pei and Fall (1995) have made the most detailed and complete analysis
of this problem. To account for the effects of dust they made the
simplifying assumption that all absorbers at the same redshift have the
same dust--to--gas ratio. By including in their treatment a term
specifying the mean metallicity, $Z(z)$, of the DLA absorbers they
successfully computed models of cosmic star formation and chemical
enrichment that account for the redshift dependence of the measured
$\Omega_{\rm HI}$, while allowing, in a self--consistent manner, for
the selection effect of the increasing obscuration due to dust as star
formation progresses. The outputs of the calculations are the evolution
with redshift of the true $\Omega_{\rm HI}$, the rate of star formation
and the metallicity of the gas.

Pei and Fall confirmed that a modest range in the dust--to--gas ratio
in DLAs at any redshift does not greatly alter their results. The
actual range in the dust--to--gas ratio is difficult to measure,
because any dusty DLAs will have been missed. Boiss\'{e} et al (1998)
provide a discussion of this issue and present evidence for a bias
against dusty DLAs in the current samples. Therefore, it appears at
least possible that the actual range in dust--to--gas ratios is large,
and that many dusty DLAs have evaded the census, which would mean that
our picture of the history of gas consumption is incorrect. A large
sample of DLAs selected from spectra of a sample of quasars
flux--limited in the K--band would provide the data for a proper
treatment of the effects of dust.

\subsubsection{Gravitational lenses and the value of the cosmological constant}

Fukugita et al (1990) and Turner (1990) show that for flat geometries
as predicted by inflation, $\Omega_{matter}+\Omega_{\Lambda}=1$, the
probability that a quasar is gravitationally lensed is more than an
order of magnitude greater for large values of the cosmological
constant $\Lambda$ ($\Omega_{\Lambda}\sim 1$) than for small values
($\Omega_{\Lambda}\sim 0$). Therefore, counting the fraction of
gravitational lenses in a sample of quasars is in principle a powerful
method for establishing the global geometry of the Universe. In
practice the calculations necessary are quite involved (e.g. Maoz and
Rix 1993, Kochanek 1996) but the conclusion of most analyses has been
that the statistics of lensed quasars and AGN are inconsistent with a
large value for the cosmological constant. For example Falco, Kochanek
and Munoz (1998) provide the $2\sigma$ limit $\Omega_{\Lambda}<0.62$.
However, in a recent study, Chiba and Yoshii (1999) found
a best fit value $\Omega_{\Lambda}\sim 0.7$. The difference between
these results is largely explained by uncertainties in the properties
of the lensing galaxies (velocity dispersions, space densities), which
probably prohibit a definitive answer until they are better
established. Another potentially significant source of uncertainty is
our lack of knowledge of the evolution of massive galaxies, in
particular their merging history.

All calculations agree that massive early--type galaxies dominate the
lensing cross--section. Locally, most early--type galaxies contain
little dust so the existence of a number of highly reddened, lensed,
quasars (e.g. MG0414+0534, Lawrence et al 1995) is something of a
puzzle. The reddening may be intrinsic to the quasar but reddening due
to dust in the lensing galaxy is also a possibility. The lensing
cross--section of massive spiral galaxies may have been
underestimated, or early--type galaxies at redshifts $0.5<z<1.0$ could
be substantially dustier than those seen locally. In either case if a
population of dusty galaxies contributes significantly to the lensing
cross section at these redshifts, then samples of gravitational lenses
drawn from optical samples of quasars would be significantly
incomplete, producing an underestimate of the value of
$\Omega_{\Lambda}$.  The uncertainty over the effect of dust on the
lensing statistics could be greatly reduced by searching for
gravitational lenses in a K--band, flux--limited sample of quasars.

\begin{figure*}
\vspace{20cm}
\includegraphics{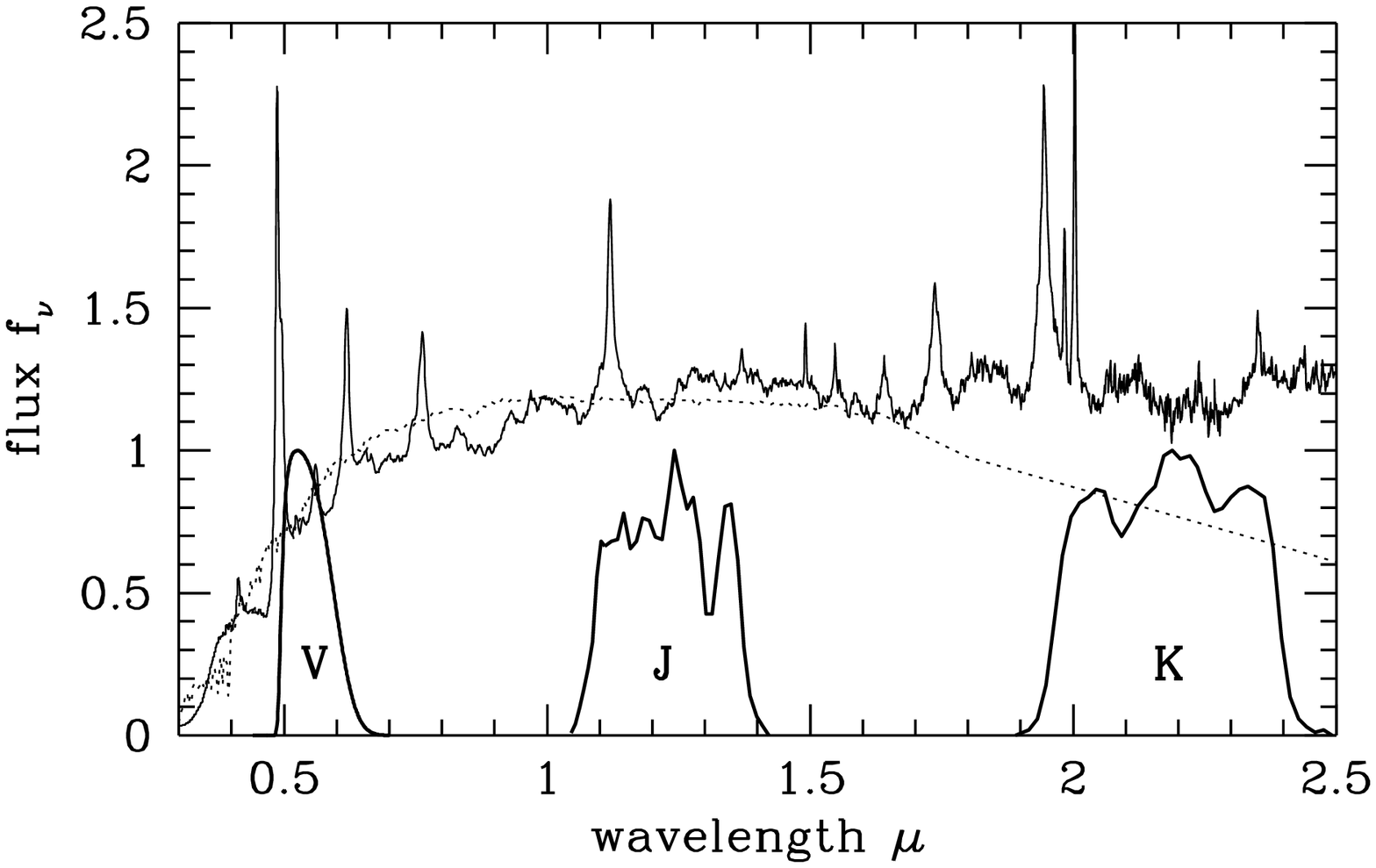}
\includegraphics{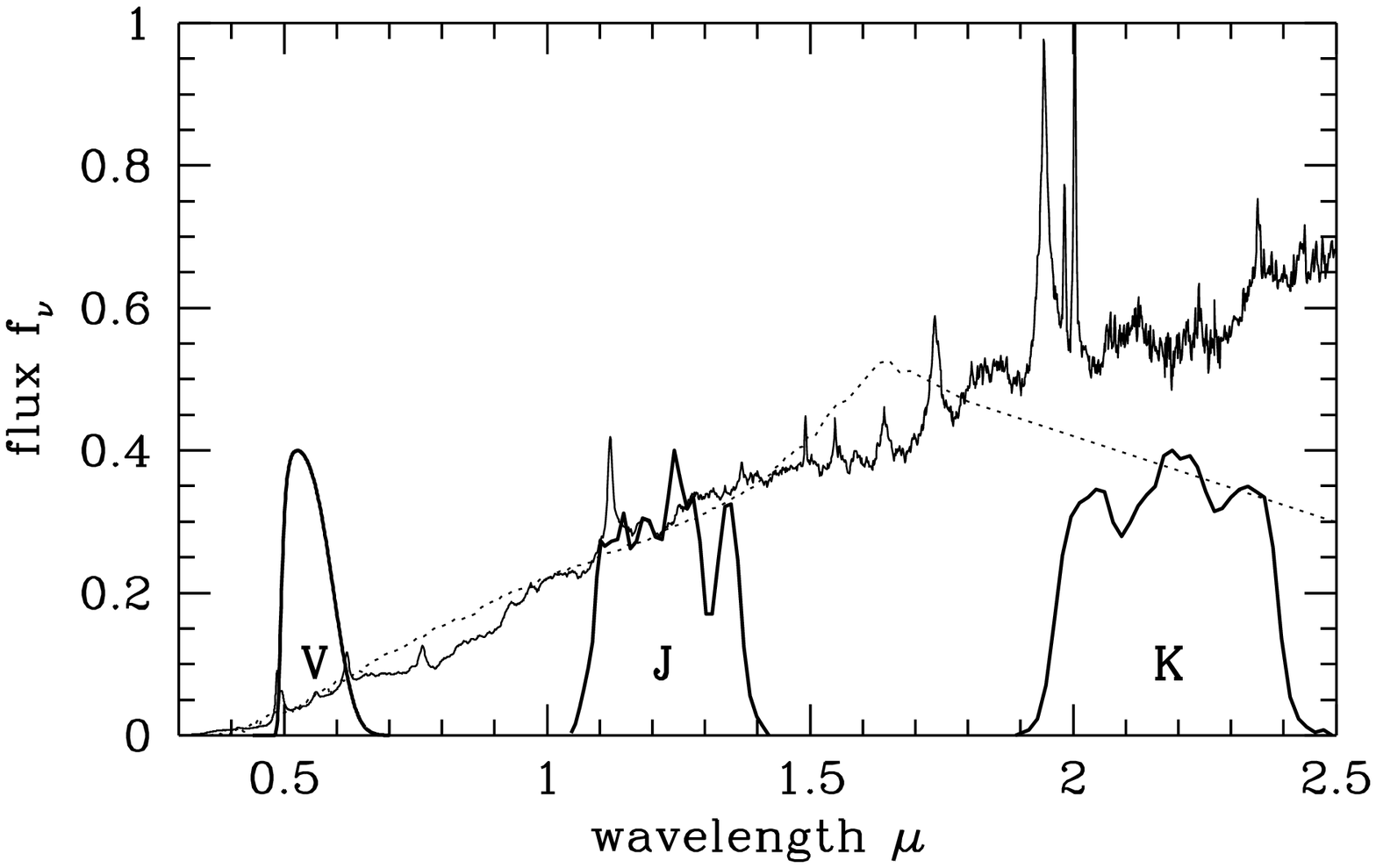}
\caption{Comparison of the optical--to-near--infrared spectral energy
distribution of quasars and stars, illustrating the principle of the KX
method. The flux scale is arbitrary, but the same for both plots. Each
plot shows an extended version of a composite quasar spectrum (Francis
et al 1991) and a star (Buser and Kurucz 1992, Kurucz 1979) with
similar V-J colour. The V, J and K filter transmission curves are
overplotted.  The upper plot shows the spectra of an unreddened $z=3$
quasar (solid line) and an early K star (dotted line). The K excess is
clearly visible. The lower plot shows the same quasar experiencing
rest--frame extinction of $E(B-V)=0.3$ in an intervening system at
$z=2.5$. The spectrum of an early M--star, having a similar V-J colour,
is shown for comparison. Again, a substantial K excess is apparent,
demonstrating the effectiveness of the KX method in identifying both
unreddened and reddened quasars.}
\end{figure*}

\subsection{Red quasars}

Quasars have been detected at all wavelengths from gamma--rays to
radio waves and they produce a significant fraction of their energy
output over many decades in frequency. Furthermore, the proportion of
the total energy radiated at different frequencies varies
substantiallly among the quasar population. A deep survey at, say,
optical wavelengths could miss quasars where the bulk of the energy is
emitted in, say, the far--infrared, and {\it vice versa}.  To
characterise the bolometric energy output of the quasar population it
is necessary to undertake surveys at several wavelengths that include
significant contributions from the different components that make up
the quasar spectrum (Hewett and Foltz, 1994).  The importance of
undertaking surveys at multiple frequencies is illustrated by Webster
et al (1995), who claim that, due to the effects of extinction, the
bulk of the quasar population has gone undetected in optical
surveys. This conclusion is nevertheless controversial (e.g.  Benn et
al 1998) and surveys for quasars in the K band would provide a direct
test of the hypothesis by establishing the fraction of reddened
quasars that are under--represented in optical samples.

\section{The KX method}

The KX method exploits the fact that, by comparison with a star with
the same V-J colour, quasars are redder in J-K so the two populations
separate in the two--colour diagram for point sources. The upper plot
in Figure 1 shows the spectrum of a quasar of redshift $z=3$ and the
V, J and K filter transmission curves. Overplotted is the spectrum of
an early K--star, chosen because its V-J colour is similar to the
quasar. The excess flux of the quasar in the K band, amounting to
$\sim 0.5\,$mag, is evident.  The lower plot in Figure 1 shows the
effectiveness of the KX method in detecting also reddened quasars. The
quasar spectrum of the upper plot has been reddened as appropriate for
an absorbing cloud at $z=2.5$ with rest--frame
$E(B-V)=0.3$. Overplotted is the spectrum of an M--star, chosen for
the match in V-J colour. The excess flux of the quasar in the K band
is again clearly visible.

The KX method in practice is illustrated in Figure 2 which is a VJK
two--colour diagram showing the location of Galactic stars ($\times$)
and quasars ($\bullet$). The stellar photometry was taken from the
list of bright UKIRT standards for which the quoted photometric errors
are $0.015$ in J-K and $0.05$ in V-J.  The VJK quasar photometry
(Hewett et al, in preparation) consists of 152 quasars, $0.2 < z <
3.4$, from the Large Bright Quasar Survey (Hewett, Foltz and Chaffee
1995) and 20 quasars, $2.0 < z < 3.4$, $16.5 < V < 19.5$, used in
studies of DLA absorbers (e.g. Pei, Fall and Bechtold 1991; Table
1). For clarity, individual error bars are not plotted but the
photometric errors are nearly all $\le 0.15\,$mag in each colour. The
V photometry was not in general acquired at the same epoch as the
infrared magnitudes and an additional scatter of $\sim 0.1\,$mag, due
to intrinsic photometric variability in the quasars, will be present
in the V-J colour for many of the objects.

\begin{figure}
\vspace{11cm}
\includegraphics{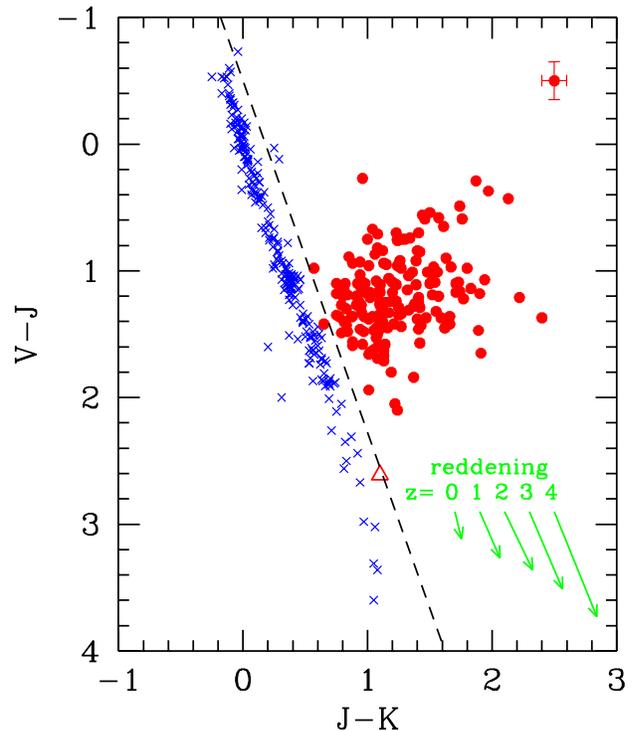} 
\caption{VJK two colour diagram for stars ($\times$) and quasars ($\bullet$)
illustrating the KX technique. The stellar photometry was taken from
the list of bright UKIRT standards at {\it
http://www.jach.hawaii.edu/UKIRT/astronomy/calib/ bright\_stds.html}.
The quasar photometry consists of our own measurements of 172 quasars,
redshifts $0.2 < z < 3.4$, magnitudes $B <\sim 19$, representative of
objects identified using optical survey techniques. Photometric errors
for the stars are similar to or smaller than the symbol size while
representative errors for the quasars are shown at top right.
Reddening vectors are shown for 5 different redshifts of the absorber
and are computed for rest--frame $E(B-V)=0.1$ using the LMC extinction
curve. The possible selection boundary, indicated by the dashed line,
is ${\mathrm J-K>0.36\,(V-J)+0.18}$. The reddening vectors are
approximately parallel to the selection boundary, so the
effectiveness of the method is largely insensitive to the effects of
reddening provided a quasar is included in the flux--limited
sample. The triangle marks a quasar of redshift $z=4.5$ discussed in
the text.}

\end{figure}

A possible selection boundary, ${\mathrm J-K}>0.36\,({\mathrm
V-J})+0.18$, discriminating the quasars from the sequence of stars is
shown by the dashed line. All but one of the 172 quasars plotted, $z <
3.4$, lie to the right of the selection line shown. The quasar to the
left of the selection line is LBQS1212+1445, a broad absorption line
object at redshift $z=1.63$. The K--photometry for this object in the
UKIRT dataset that provides many of the infrared magnitudes shown in
Figure 1 is $0.4\,$mag fainter than an observation made with the
Multiple Mirror Telescope (MMT). While the use of the MMT K--magnitude
would move the object $0.4\,$ mag rightward into the well populated
portion of the plot there is no indication of anything amiss with the
UKIRT K-band magnitude and for consistency we have plotted the J-K
colour from the UKIRT observations.

Also shown In Fig. 2 are the reddening vectors for an intervening
absorber with an LMC extinction curve, of rest--frame $E(B-V)=0.1$, at
five different redshifts. Each vector runs approximately parallel to
the stellar locus, so reddened quasars can also be detected by this
method. \footnote{We checked that this conclusion also holds true for
the UV extinction curves for the SMC (Pr\'evot et al. 1984) and for
the Galaxy (Mathis 1990). In the case of the SMC at high redshift the
reddening vectors are a little steeper than those for the LMC, while
for the Galaxy the vectors are a little shallower.}  Given adequate
signal--to--noise ratio, $\sim 10$, in the estimate of the object
colours the KX method should identify nearly all quasars found in
optical samples, whether reddened or not, above the K--band flux
limit.

Application of the KX method in practice will require good quality
images in order to separate point sources from galaxies, because the
colour distribution of faint galaxies overlaps that of quasars (this
is also true of the UVX method).  A potential difficulty which we have
not addressed is the brightness of the quasar host galaxy. The
contribution to the total flux of the host galaxy will in general be
larger in the K band than in the optical so lower luminosity quasars
might be excluded from a catalogue of point sources. However,
this is not an issue for the bright flux levels considered in \S3.

The VJK combination is so effective because the spectral energy
distribution of stars with similar V-J colours to quasars (principally
those of spectral types K and M) turn over in the H band (Fig. 1). The
J and K bands straddle the break point, providing the discriminatory
power of the technique, so utilising H--band magnitudes in place of J
or K is not viable. A limit to the effectiveness of the KX method using
V, J, and K passbands occurs at redshifts $z>3.5$ where absorption by
the Ly$\alpha$ forest is present over much of the wavelength range
included in the V--filter.  Quasars start to become redder in V-J,
moving vertically in the VJK diagram and approaching the stellar
locus.  In Figure 2 the triangle marks the colours of a quasar of
redshift $z=4.5$, which lies below the colour selection boundary
because of this absorption. Substituting the R or I filter for V would
extend the effectiveness of the KX method to redshifts beyond $z=4$.

At the end of this section we reemphasise the difference between the
effects of reddening and extinction on the completeness of quasar
surveys. Some optical survey methods, including the multicolour method
and emission-line searches (but excluding the UVX method), are also
largely insensitive to reddening. The advantage of the KX method over
all optical survey methods is the reduced extinction in the K band,
rather than simply the ability to find reddened quasars, i.e. a much
larger fraction of quasars suffering extinction will be included in
the K--band, flux--limited sample.

\section{A KX survey with the UKIRT wide-field near-infrared camera}

The proposed UKIRT wide--field near--infrared camera will image 0.2
square degrees per exposure. One of its goals is a moderately deep
survey over a substantial fraction of the area of the Sloan Digital Sky
Survey (SDSS) accessible by UKIRT, i.e. thousands of square degrees,
to a depth of K=19 with a signal--to--noise ratio of $\sim 10$. Such a
survey would contain many thousands of quasars. Here we consider the
effectiveness of this survey for finding damped Ly$\alpha$ galaxies and
gravitational lenses, especially examples where the background quasar
had been dimmed by dust in the intervening galaxy.

\subsection{Damped Ly$\alpha$ galaxies}

A survey for DLAs requires samples of high--redshift $z \ga 2$
quasars, since for ground--based spectroscopy DLAs are only detectable
at $z>1.8$. The number of DLAs catalogued in the literature is
approaching 100. Therefore, to provide a substantial advance, a survey
that produces more than 300 DLAs is desirable. For a given survey area
we can compute the K-band flux limit that will provide sufficient
quasars to produce a sample of DLAs of this size. We can then verify
that this flux limit is sufficiently bright, say R$\le20$, to allow
high--resolution spectroscopy on an 8--metre class telescope for the
measurement of the absorber metallicities.

For the calculation we take an area of 4000 square degrees over a
region in common with the SDSS. An advantage of covering the SDSS area
is that low--resolution spectra of most of the bright KX--selected
quasar candidates will exist in the SDSS database. Taking the quasar
luminosity function of Warren, Hewett and Osmer (1994) we can compute
the surface density of high--redshift quasars as a function of R
magnitude. We then convert to K assuming a mean colour of R-K=2.2. We
then estimate the number of DLAs by assuming a line density
$dn/dz=0.055(1+z)^{1.15}$ (Wolfe et al. 1995) and supposing that DLAs
can be detected from the quasar emission redshift down to $z=1.8$. 

Over 4000 square degrees, to K=16.0, we estimate there would be $\sim 2300$
high--redshift, $z>2.2$, quasars with over 500 detectable DLAs.  If
there is a serious bias in existing optical samples of DLAs the number
of DLAs detected will be larger. Since K=16.0 corresponds to R=18.2
for a typical quasar, quasars that are substantially reddened, $\la
2\,$mag in the optical, would still be bright enough for
high--resolution spectroscopic follow--up.

An alternative strategy could be to use radio--selected quasars to
undertake a survey for DLAs unbiased by extinction due to dust. There
are two disadvantages to this approach. Firstly there are insufficient
high--redshift radio quasars to yield a sample of DLAs of the size
envisaged here. Secondly, a significant fraction of the quasars will
be very faint in the optical, precluding high--resolution
spectroscopy, yet it is essential to survey the spectra of all the
quasars for the sample to be unbiased. Nevertheless, any subset of
quasars suffering very large extinctions that still elude the K--band
flux--limited selection could be identified in a radio survey.

\subsection{Gravitational lenses}

Excluding lensing by galaxy clusters, there are currently only some 40
examples of strong gravitational lensing known ({\it
http://cfa--www.harvard.edu/castles/}). For the same survey for bright
quasars considered above, $K<16.0$, counting quasars of all redshifts
there will be over 10000 quasars over the 4000 deg$^2$ survey
area. For a typical quasar colour V-K=2.5 the survey magnitude limit
is equivalent to V=18.5. Therefore the apparent magnitudes of the
quasars will be similar to the apparent magnitudes of the quasars
observed in the HST snapshot survey (Maoz et al 1993) which have
$V=18.0\pm0.8$. There are five cases of gravitational lensing amongst
the 502 quasars imaged in the snapshot survey.  Because the apparent
magnitudes are similar we can use the results of the snapshot survey
to estimate the number of lenses in the UKIRT survey by assuming that
$1\%$ of the quasars are lensed (the fraction of lenses in a sample
depends on the sample depth because of magnification bias). Therefore
the UKIRT survey should produce roughly 100 gravitational lenses. The
number of lenses detected could be larger than 100 if bias due to dust
is important. 

It is unlikely that the image quality achieved in the survey will be
good enough to detect examples of gravitational lensing where the
image separation is less than $0.5\,$arcsec. Sensitivity to
separations as small as $0.2\,$arcsec could be achieved by later
imaging all the bright quasars using adaptive optics. The fraction of
lenses in a K-selected sample could be increased in two ways. Firstly,
a sample of similar size could be obtained by surveying a larger area
of sky but to a brighter magnitude limit, thereby increasing the
magnification bias.  Alternatively, follow--up high--resolution
imaging could be limited to quasars of high redshift, since the
probability that a quasar is lensed increases with redshift.

The interpretation of a survey for gravitational lenses requires
knowledge of the quasar luminosity function at magnitudes fainter than
the search limit, since the lensed quasars have been magnified. A
potential drawback of the KX method is contamination of the sample of
candidate quasars at faint magnitudes by compact galaxies that
morphologically cannot be distinguished from stars. Nevertheless, at
these fainter magnitudes where the surface densities are higher it will
be feasible to measure the quasar luminosity function based on
spectroscopic surveys of complete flux--limited samples (i.e. with no
colour or morphological selection) using multi--object spectrographs.
One advantage of K--band surveys for gravitational lenses over radio
surveys such as CLASS (Jackson et al 1998) is the ease with which the
redshift distribution of the unlensed source population can be
measured.

\subsection*{Acknowledgements} We thank Scott Croom for comments on the
draft. The authors acknowledge the data and analysis facilities
provided by the Starlink Project which is run by CCLRC on behalf of
PPARC.

\end{document}